\documentclass[preprint]{aastex}
\begin{document}
\title{Mapping the X-ray Emitting Ejecta in Cassiopeia A with Chandra}
\author{Una Hwang (1,2), Stephen S. Holt (1), \& Robert Petre (1)}
\affil{(1) NASA Goddard Space Flight Center, Greenbelt, MD 20771, \\
(2) Department of Astronomy, University of Maryland, College Park, MD
20742}

\begin{abstract}
We present X-ray emission line equivalent width images of the bright
Galactic supernova remnant Cassiopeia A for the elements Si, S, Ar,
Ca, and Fe using a 50,000 s observation with the Advanced CCD Imaging
Spectrometer on the Chandra X-ray Observatory.  The images essentially
identify the bulk of detectable ejecta of these elements over a wide
range of surface brightness, and show morphologies distinctly
different from that of the broadband X-ray emission and of the 4-6 keV
continuum emission.  The Si, S, Ar, and Ca maps, while different in
turn from those of Fe, show that these X-ray emitting ejecta are
distributed similarly to the fast optical ejecta knots, and clearly
delineate the X-ray counterpart of the northeast optical jet.  Low
surface brightness regions just outside the bright shell in the north
and west are also shown to have strong line emission.  The strong Fe
emission is exterior to that of other elements in the east, as
previously noted, but is generally coincident elsewhere.  The
projected interior has relatively little emission traced by high line
equivalent widths.
\end{abstract}

\section{Introduction}

A record of a supernova explosion is imprinted in the spatial
distribution of ejecta in the remnant, but then is gradually erased
by dynamic interaction with the surrounding medium.  Cassiopeia A, the
brightest and youngest known Galactic remnant, is one of the best
targets for study of the ejecta distribution.  Detailed observations
and modelling at optical wavelengths (e.g., Reed et al. 1995, Lawrence
et al. 1995, Fesen \& Gunderson 1996) show a complex system of
chemically enriched knots with high velocities that form a shell with
a linear extension to the northeast called the jet. Although these
ejecta are nearly undecelerated, they show asymmetries in their red-
and blue-shifted velocities and in their line intensity ratios that
arise from their interaction with an inhomogeneous medium.

Optical emission, however, accounts for only a small fraction of the
total mass, as typical shocks in young supernova remnants heat the
bulk of the mass to X-ray emitting temperatures.  X-ray spectral
imaging, which is required to identify and map these ejecta, has 
been carried out most extensively with the ASCA Observatory (e.g.,
Holt et al. 1994, Fujimoto et al. 1995, Hwang \& Gotthelf 1997; also
Vink et al. 1999 with SAX).  The moderate spectral resolution of the
ASCA CCD detectors allows strong line blends of individual elements to
be imaged with the 3$'$ half-power diameter point spread function of
the ASCA mirrors.  With the launch of the Chandra X-ray Observatory
comes a tremendous increase in the capability to make these
observations.  The Chandra mirrors provide $<0.5''$ imaging
capability, which is on par with excellent ground-based optical
observations, while the Advanced CCD Imaging Spectrometer (ACIS) on
Chandra gives spectral resolution comparable to that of the ASCA CCDs.

Cas A was the Chandra first light target in 1999 August, and Hughes et
al. (2000) use this early observation to note the existence of
Fe-dominated zones in the east that are exterior to Si-dominated
zones.  Because Fe is synthesized in the innermost layers of the star,
this is interpreted as an overturning of ejecta layers in this region
during the explosion.

In this paper, we present the global distribution of the X-ray line
emission in Cas A using a new Chandra observation that is ten times
longer than any previous one with ACIS.  We account for the locally
variable underlying continuum to produce maps of the line-to-continuum
ratio that reveal the distribution of strong line emission throughout
the remnant over a wide range of surface brightness.  These are the
first such images that have been produced for X-ray emission from all
these elements in any young supernova remnant.

\section{Emission Line Images}

Cas A was observed on 2000 January 30-31 using the backside
illuminated ACIS chip S3 for 50,000 s in graded mode (i.e., each event
is assigned a grade on board the spacecraft, rather than in later
processing, that is based on the local pixel intensity distribution in
the 3.24 s readout cycle). Only the standard ASCA grade 02346 events
are retained.  The light curve is stable, and in any case the source
count rate is much higher than the background contribution.  Some 16
million photons were collected during the observation, which was
carried out at a CCD operating temperature of -120 C.

The ACIS spectrum obtained for the entire remnant is shown in Fig. 1.
It features prominent He~$\alpha$ (n=2$\rightarrow$n=1) blends of Si,
S, Ar, and Ca, plus emission from L and K transitions of Fe, and a
weak Mg blend that is further blended with Fe L emission.  The
spectrum at lower energies is strongly attenuated by the high column
density of neutral hydrogen towards Cas A.  The ACIS spectrum is
comparable to that obtained by ASCA (e.g., Holt et al. 1994), except
that the ACIS S3 blends the weak Ly~$\alpha$ lines with the stronger
He~$\alpha$ emission.

We selected photons corresponding to a particular spectral feature by
identifying appropriate pulse height ranges, as indicated in Fig. 1.
Except for Fe, we combine the He~$\alpha$ and Ly~$\alpha$ emission of
each element in a single image.  We do not include Mg in this paper
because it is both blended with Fe L and highly absorbed.  We have
made no correction for the spatially varying gain across the CCD, as
this effect is small compared to the width of the pulse height ranges
used.

In mapping the line emission, we correct for the underlying continuum,
as the continuum may well have a different spatial distribution than
the line emission. In a solar abundance plasma, the continua are
associated primarily with lower atomic weight atoms and their
electrons, and may also include a contribution from a highly
energetic, nonthermal population of electrons.  Indeed, the X-ray
image of Cas A in the line free spectral region of energies between
4-6 keV more closely resembles radio images than it does the
line-dominated X-ray broadband image (see Fig. 2, also Holt et
al. 1994, Vink et al. 1999, Koralesky et al. 2000, Gotthelf et
al. 2000).  Because X-ray line and thermal continuum emissivities both
scale with the emission measure, the line emission in a region of low
surface brightness may still be strong relative to the continuum.  To
identify ejecta over a wide range of surface brightness, we therefore
subtract the continuum from each line image and then form a ratio
image relative to the continuum (hereafter referred to loosely as a
line equivalent width image, EQW, for brevity).  A ratio image of much
lower spatial resolution was presented for the Fe K blend in Cas A by
Vink et al. (1999).

An optimal continuum subtraction should account for the local shape of
the spectrum for a source like Cas A, where the spectrum changes on
small spatial scales (Hughes et al. 2000).  To do this efficiently, we
extracted images in two pulse height bands on either side of that used
for the line image (see Fig. 1).  These bands minimize the
contamination from weak line emission and are in some cases quite
narrow, but generally contain over 100,000 counts.  The continuum is
assumed to be linear throughout the range of pulse heights used for a
particular image, and is interpolated pixel by pixel using 1$''$
pixels for the Si, S, and Fe L images, and 2$''$ for the others.  The
continuum images for Fe K are sparse, so in this case only, they were
simply added and scaled to the appropriate average level inferred from
fitting the total spectrum.  In the faint regions interior and
exterior to the bright broadband shell, some pixels are negative after
continuum subtraction, so each raw line and background image is
smoothed with a 1-2 pixel gaussian filter before subtraction and
division.  The final line and line/continuum ratio images are smoothed
over 1.5 pixels.


The EQW images for Si, S, Ar, Ca, and Fe (L and K) are shown in Fig.
3.  The Si, S, Ar, and Ca images are, on the whole, similar to each
other, but notably different from the broadband image and the 4-6 keV
continuum.  Comparison of the Si EQW image to the intermediate,
continuum subtracted Si line image in Fig. 2 shows that the effect of
normalizing to the continuum is important. The optical jet to the
northeast is strikingly prominent in these EQW images, having EQW
intensities comparable to those seen at the bright ejecta shell.
Using the Si EQW image to guide the eye, the distribution of Si
emission with high EQW is seen to strongly resemble the distribution
of fast-moving optical [SII] ejecta knots (e.g., Fesen \& Gunderson
1996), including the ring-like structure noted by Lawrence et
al. (1995) in the northern arch.  It is interesting that the regions
just exterior to the bright shell to the north and west in the
broadband and Si line images also show high EQW intensities (compare
the Si images in Fig. 2 and 3 against the fiducial marker), whereas
low EQW Si emission is seen throughout the remnant in regions off the
bright shell.

The remnant's eastern boundary in the EQW images of the intermediate
mass elements appears more prominent than in the optical, and is
closer in to the center of the remnant than in the broadband X-ray
image.  The remaining panels of Fig. 3 show that the Fe L and Fe K
maps are similar to each other in that their eastern boundaries are
exterior to that of the other elements, as previously noted (Hughes et
al. 2000, Vink et al. 1999).  In addition, Fe emission is weak or
absent in the jet region and is contained within the boundary of the
bright emission to the north and west seen in other images.  We expect a
general similarity between the Fe L and K images since Fe K emission
is usually accompanied by L emission, but the converse need not be
true, and there are regions that are bright in Fe L but not in Fe K;
presumably the regions lacking Fe K emission are at lower
temperatures.  The deficiency of Fe L in the west can be largely
attributed to the higher neutral H column density toward this part of
the remnant (Keohane et al. 1996), and this also accounts for the
weaker Si emission in this region.  A final, but important, caveat is
that the emission we have identified as ``Fe L'' may well include a
contribution from Ne, an issue to be fully resolved with data of
higher spectral resolution.

\section{Discussion}

It can be verified that the maps in Fig. 3 trace gas that is enriched
with supernova ejecta by comparing the observed line intensity ratios
to theoretical line/continuum emissivity ratios.  Using the
approximate Si energy range shown in Fig. 1 for a solar abundance
plasma with temperatures between $10^6-10^8$ K and ionization ages
from $10^9-10^{12.4}$ cm$^{-3}$ s, the calculated ratio of the Si line
emissivity relative to the continuum does not exceed 2 (and this only
at extreme temperatures).  A ratio of 2 traces the bright regions of
the Si EQW image in Fig. 3, which is consistent with the consensus
that the X-ray Si emission in Cas A comes predominantly from ejecta.

A similar calculation for Fe K gives maximum ratios of just over 2 for
the same range of parameters, with the highest computed ratios
occurring at temperatures of a few keV near collisional ionization
equilibrium.  A ratio of 2 again outlines the Fe K EQW image in
Fig. 3, suggesting that much of this emission may be from ejecta.
Another possibility has been suggested by Borkowski et al. (1996), who
modelled the Fe K emission in the integrated ASCA spectrum as arising
from shocked circumstellar wind material.  The images do not rule out
the possibility that some of the Fe K emission does come from such a
contribution, and to some extent, this could also be true of the
emission from Si and other elements.  The resolution of this issue
requires detailed examination of the spectra, which we defer to future
work.

The images in Fig. 3 give a good indication of the distribution of
X-ray emitting ejecta, but we emphasize that they do not give detailed
abundance information.  Theoretical equivalent widths vary with
temperature and ionization age, so that a given equivalent width can
correspond to significantly different element abundances depending on
the local plasma conditions.  The EQW maps by themselves should
therefore be taken only as an indication of where the abundances are
definitely enhanced above the solar value.  The details of the maps,
particularly in faint emission regions, are moreover dependent on the
rather simple assumptions that the local continuum is linear in shape
and free of contamination from weak lines, and that the line emission in
each image is due entirely to the element identified.  Though this
last assumption is likely to be quite valid in most cases, it may not
be valid for Fe L, nor would it be for Mg.  Finally, the EQW maps by
themselves are not actual maps of the ejecta mass, as the mass depends
not only on the abundance and emitting volume but also on the local
density.

The resemblance between our Si and S line maps and optical maps
explicitly demonstrates that the X-ray and optically emitting ejecta
are spatially coexistent over much of the remnant.  This was
anticipated in that the X-ray Si and S line emission, inferred to
include a significant contribution from ejecta, show asymmetric
Doppler shifts corresponding to bulk velocities of $\sim$2000 km/s
that are comparable to those of the optical ejecta knots (Markert et
al. 1983, Holt et al. 1994).  The ejecta in a given region therefore
have a wide range of densities and temperatures, and the shocks
propagating into the ejecta are highly variable on small scales; the
slow shocks required for the optical emission (Reed et al. 1995)
cannot explain the X-ray emission.

The faint X-ray jet extends more than 3.5$'$ beyond the expansion
center used by Kamper \& van den Bergh (1976, a position nearly
coincident with that of the newly discovered X-ray point source)
making it of comparable extent to its optical counterpart.  The
qualitative resemblance between the X-ray image of the jet and the
optical images of Fesen \& Gunderson (1996) is striking.  In
particular, the three prominent filaments of the outer optical jet and
the triangular shaped base are all seen to have X-ray counterparts.
Our maps also show that the abundances of Si and S are enhanced in
faint regions just outside the bright shell in the north and west, but
they do not reveal any other extended region of enhanced abundance
comparable to the northeast jet.  The Si image, containing the highest
number of photons, does show faint streamers of emission extending
radially outward directly opposite the jet to the west, but with a
larger opening angle, and with equivalent widths that are not high
enough to indicate unambiguously that these are ejecta.

We have noted that Si, S, Ar, and Ca have similar general
morphologies, but examination of the individual panels of Fig. 3 does
show differences between them.  A detailed survey of such differences
is beyond the scope of this paper, but two notable examples are the
brightest knots in the Si image of Fig. 2, at the inner boundary of
the shell to the northeast and southeast: these are much weaker in the
Ar and Ca EQW maps than in Si and S.  The southeast knot was analyzed
qualitatively by Hughes et al. (2000), and it appears that the Ar and
Ca abundances here are indeed lower, and that these ejecta originate
in an outer ejecta zone compared to knots with strong emission for all
four elements.

The overall difference between the Si and Fe EQW maps in the east
appears to support the suggestion that the inner Fe ejecta layers have
been overturned and propelled beyond the Si ejecta layers in this part
of the remnant, and may be related to the asymmetries that created the
jet nearby.  Such overturning of ejecta layers was first suggested by
Chevalier \& Kirshner (1979 and references therein) to explain the
properties of the optical knots, and also occurs in recent models
(e.g., Kifonidis et al. 2000, Khokhlov et al. 1999).

In summary, the X-ray line maps presented in this paper reveal the
spatial striking correspondence between the optically and X-ray
emitting ejecta in Cas A, explicitly showing that the different
physical conditions giving rise to each coexist on small spatial
scales.

\acknowledgments
We thank Paul Plucinsky for expert advice during the planning of this
observation, and Harvey Tananbaum for gracious assistance in
scheduling it.

\begin{figure}
\centerline{\includegraphics[scale=0.7]{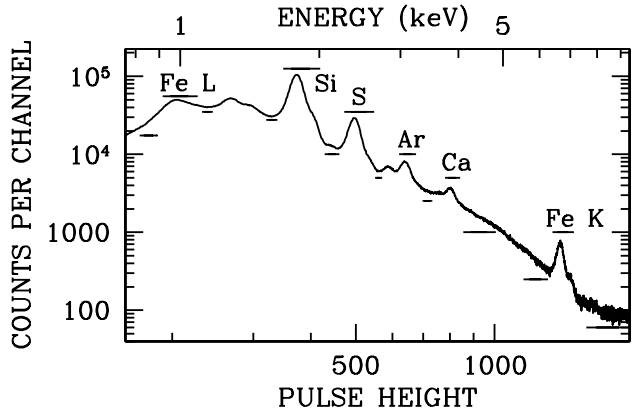}}
\vspace{-3.0in} 
\figcaption{The ACIS spectrum of Cas A plotted against both pulse
height and energy (using an approximate gain value of 4.8 eV per
channel).  Horizontal bars above the spectrum show the pulse height
intervals used for each image, also labelled by element; bars below
the spectrum show the neighboring intervals used for the continua.}
\end{figure}

\newpage
\begin{figure}
\centerline{\includegraphics{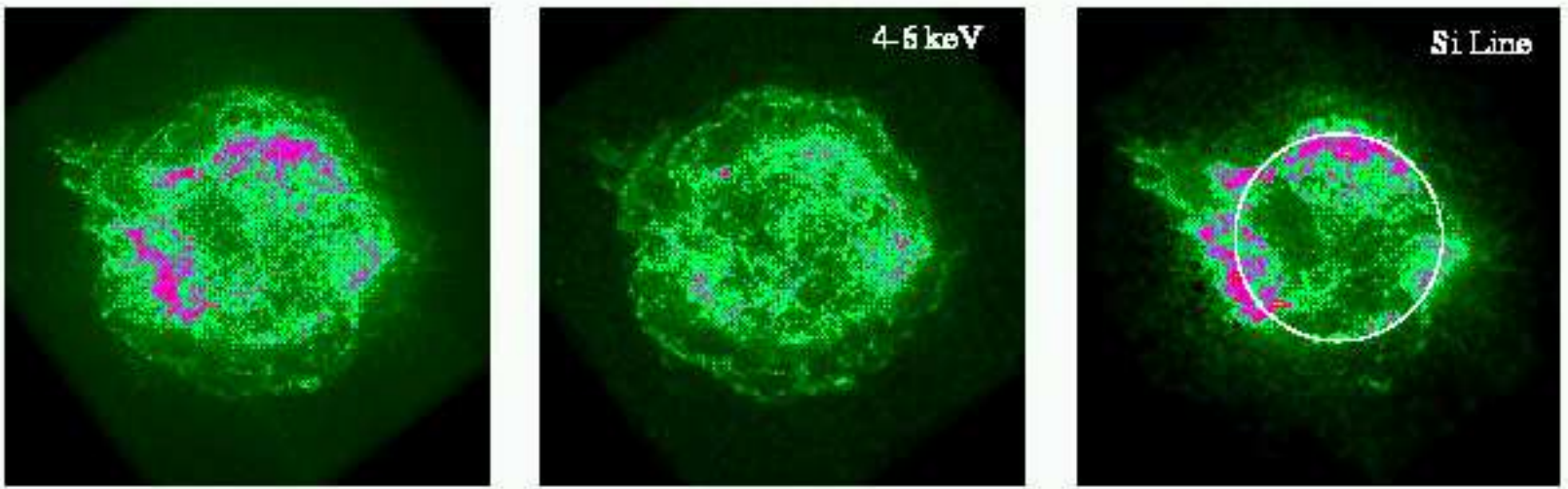}}
\figcaption{
(Left) The broadband ACIS S3 image of Cas A.  (Middle)
The 4-6 keV continuum image. (Right) The continuum-subtracted Si line
image with a fiducial circle of radius 1.8 arcmin overlaid. All
images are 512 arcsec across with square-root intensity scales.}
\end{figure}
\begin{figure}
\centerline{\includegraphics{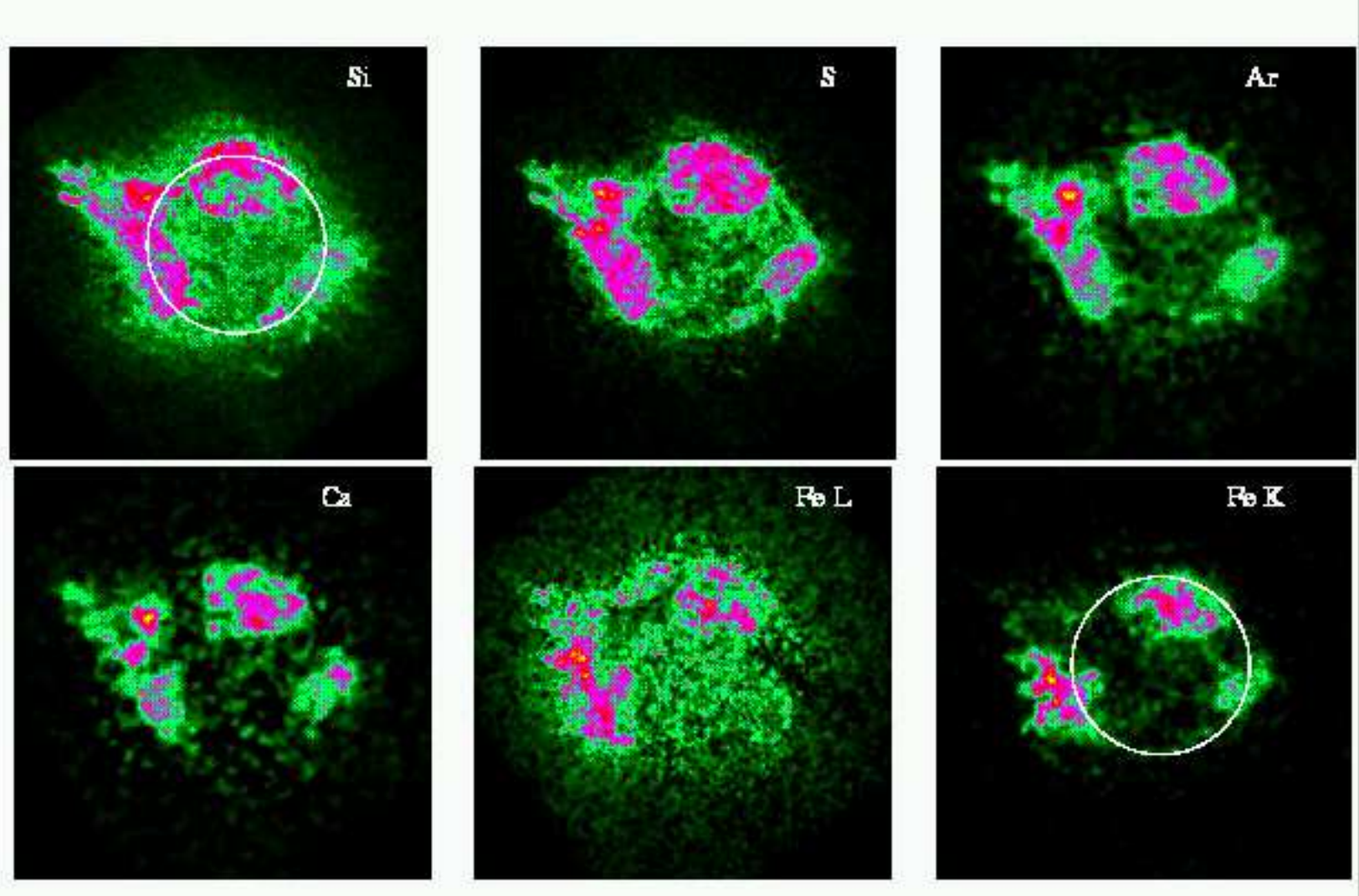}}
\figcaption{ Left to right: Line-to-continuum ratio (equivalent width)
images for (top) Si, S, Ar, (bottom) Ca, Fe L, and Fe K.  The same
fiducial circle in Fig. 2 is overlaid on the Si and Fe K images to
facilitate comparisons; the intensity scales are linear and negative
ratios have been set to zero.}
\end{figure} 

\end{document}